\documentclass[preprint]{elsarticle}
\usepackage[utf8]{inputenc}
\usepackage[a4paper, margin=1in]{geometry}
\usepackage{amsmath}
\usepackage{amssymb}
\usepackage{color}
\usepackage{subcaption}
\usepackage{graphicx}
\newcommand{\R}{\mathcal{R}}
\newcommand{\SA}{\mathcal{S}}
\DeclareMathOperator*{\argmin}{arg\,min}
\usepackage{amsthm}
\newtheorem{theorem}{Theorem}[subsection]
\newtheorem{corollary}{Corollary}[theorem]
\newtheorem{lemma}{Lemma}[subsection]
\theoremstyle{definition}
\newtheorem{definition}{Definition}
\theoremstyle{definition}
\newtheorem{condition}{Condition}
\begin{document}
    
\begin{frontmatter}

\title{Prospect-theoretic Q-learning\tnoteref{label1}}
 \tnotetext[label1]{The work of VSB was partially supported by  the S.\ S.\ Bhatnagar Fellowship from the Government of India.}

\author[EE]{Vivek S. Borkar, Siddharth Chandak}
\ead{borkar.vs@gmail.com, chandak1299@gmail.com}

\address[EE]{Department of Electrical Engineering, Indian Institute of Technology Bombay, Powai, Mumbai-400076, India }

%
%
%
%
%

\begin{abstract}  We consider a prospect theoretic version of the classical Q-learning algorithm for discounted reward Markov decision processes, wherein the controller perceives a distorted and noisy future reward, modeled by a nonlinearity that accentuates gains and under-represents losses relative to a reference point. We analyze the asymptotic behavior of the scheme by analyzing its limiting differential equation and using the theory of monotone dynamical systems to infer its asymptotic behavior. Specifically, we show convergence to equilibria, and establish some qualitative facts about the equilibria themselves.
\end{abstract}

\begin{keyword} Q-learning; prospect theory; cooperative o.d.e.; monotone dynamics; stable equilibria
\end{keyword}

\end{frontmatter}

\section{Introduction}
Traditional reinforcement learning schemes are concerned with the actions of  rational agents seeking to maximize their expected rewards. While these rational agents are risk neutral, reinforcement learning has also been  studied under risk-sensitive (risk averse) policies \cite{Mihatsch, ShenS, ShenT}. But according to prospect theory \cite{KT} and its sibling \textit{cumulative prospect theory} \cite{TK}, human beings perceive risk differently in different scenarios: they can be risk seeking in some situations and risk averse in others. (See also \cite{Dacey, Armstrong}.) In this work, we study  classical Q-learning for Markov decision processes \cite{Watkins}, one of the early  reinforcement learning algorithms, from a prospect theoretic viewpoint, i.e., when the future returns are distorted using an `S-shaped' valuation map that increases perceived gains and decreases perceived losses. 

Previous works \cite{ShenS, ShenT} applying such prospect theoretic valuation maps worked with certain restrictive assumptions. For example, \cite{ShenS} does not allow for steep valuation maps and high discount factors for  future rewards. On the other hand, \cite{ShenT} changes the original Q-learning scheme in a manner that ensures convergence, but the formulation is a departure from the original paradigm as we point out later. In this work, we study the asymptotic behavior of Q-learning scheme when these additional restrictions and/or modifications are dropped. Naturally, we lose global convergence to a single equilibrium, but nevertheless we are able to characterize the asymptotic behavior in qualitative terms to a significant extent. The tools we use are the o.d.e.\ (for `Ordinary Differential Equations') approach to  stochastic approximation \cite{DerFrad, Ljung} (See \cite{Borkar} for a textbook treatment) and monotone dynamical systems \cite{Hirsch1, Hirsch2,Smith} to show convergence of the iteration to the set of equilibria and the structure of the latter set.  

Our motivation for this study is twofold. The first is the classical motivation behind learning models in economics, viz., to build simple dynamic models to study the qualitative behaviour of boundedly rational macroeconomic agents \cite{Fudenberg}. Some of the interesting insights we obtain are that the learning in fact equilibrates (i.e., does not get into more complicated behavior such as cycling or worse, strange attractors), though not to a unique equilibrium. Furthermore, the choice of equilibrium depends on the initial condition which is characterizable  at least in the cases when they are too high or too low in a certain sense. The equilibration of our model also has a flavor of rational expectations equilibrium \cite{Begg}, which may be an interesting analogy to pursue further. 

Secondly, with algorithms taking over from humans in many spheres of human activity, there are `human in the loop' scenarios such as e-commerce, crowdsourced decision making, recommendation networks, etc., where the empirically validated aspects of human judgement such as prospect theory must be factored in. This is so both when not doing so will lead to erroneous predictions and \textit{ipso facto} erroneous decisions, and when the correct outcomes of human peculiarities can lead to undesired outcomes and you want to correct for them. For this, a good theoretical groundwork in terms of mathematical models of behavioral dynamics are important.

The need for this is already felt in the rapidly increasing literature that tries to factor in extending prospect theoretic aspects, e.g., in finance \cite{He, Zhang}, game theory \cite{Phade1,Phade2, Phade3,Hota,Tian}, the newsvendor problem \cite{Vipin, Nagarajan, Shen-news, Surti}, energy exchange mechanisms \cite{Mandayam}, etc., the works in prospect theoretic reinforcement learning are relatively few.  In addition to those mentioned above, one interesting effort, albeit not in Markov decision theoretic framework, is \cite{LAP}. See also \cite{Denrell, Feldmaier, Ratliff} for other interesting takes on this theme. See \cite{Wakker} and \cite{Bertsekas} for  textbook treatments of prospect theory and reinforcement learning resp., where further pointers to literature in the respective fields can be found.  

In the next section, we give a short introduction to Q-learning and briefly, the aspects of prospect theory relevant here. In Section \ref{sec:scheme_convergence}, we first present the modified Q-learning scheme and then study its limiting o.d.e. We then use results from monotone dynamical systems to show its convergence to equilibrium points. Then, in Section \ref{eq.points}, we show results regarding stability, location and the number of equilibrium points for this scheme. In Section \ref{sec:simu}, we give a summary of the numerical simulations and comment on our observations. In Section \ref{sec:alt}, we discuss an alternative modification of the prospect theoretic Q-learning iteration where the total reward is prospect theoretically distorted, not only the future returns.

\textbf{Notation:} For ease of reference, we list the key notation used in the paper in the following table.
\begin{center}
\begin{tabular}{| c | c |}
\hline
 \multicolumn{2}{|c|}{Key Notation} \\
 \hline
 $\{X_n\}$ & Controlled Markov chain \\ 
 $\{U_n\}$ & Control process \\  
 $S$ & Finite state space with cardinality $s$ \\
 $A$ & Finite action space with cardinality $r$ \\
 $\alpha$ & Discount factor \\
 $k(i,v)$ & Reward for choosing action $v$ at state $i$ \\
 $k_{min}, k_{max}$ & min, resp.\ max of $k(\cdot,\cdot)$ \\
 $K$ & The constant $\frac{k_{max}}{1 - \alpha}$ \\
 $V(i)$ & Value function (see (\ref{DP})) \\ 
 $Q(i,v)$ & Q-values (see (\ref{Q-DP})) \\
 $F(\cdot)$ & Dynamic programming operator for $Q(\cdot,\cdot)$ \\
 $a(n)$ & Stepsize sequence \\
 $\mu(i,v,n)$ & Number of times action $v$ is chosen at \\
 & state $i$ till iteration $n$\\
 $u(\cdot)$ & S-shaped map (see Figure \ref{fig:u_example}) \\
 $\xi(i,v)$ & i.i.d. noise concentrated in $[-c,c]$ \\
 $\epsilon$ & Parameter for epsilon-greedy policy (see (\ref{epsilon-greedy})) \\
 $\Lambda(t)$ & Diagonal matrix for time scaling under\\
 & asynchrony \\
\hline
\end{tabular}
\end{center}

\section{Background}
\subsection{Q-Learning}

We sketch the derivation of Watkins' Q-learning algorithm \cite{Watkins} for discounted reward Markov decision processes, as a backdrop for our work. Consider a controlled Markov chain $\{X_n\}$ on a finite state space $S, |S| = s$, governed by a control process $\{U_n\}$ in a finite action space $A, |A| = r$, with controlled transition kernel $(i, j, v) \in S^2\times A \mapsto p(j|i,v) \in [0, 1]$ with $\sum_jp(j|i,v) = 1$. The controlled Markov property is
$$P(X_{n+1} = i |X_m, U_m, m \leq n) = p(i |X_n, U_n) \ \forall n.$$
Let $\alpha \in (0, 1)$ be a discount factor and $(i,v) \in S\times A \mapsto k(i,v) \in [0, \infty)$ the per stage reward. For future reference, we denote by $k_{min}, k_{max}$ the minimum and maximum values of $k(i,v)$, which we assume are distinct (i.e., $k$ is not a constant). The infinite horizon discounted reward problem is to maximize 
$$E\left[\sum_n\alpha^nk(X_n, U_n)\right]$$
over all $\{U_n\}$ as above, called `admissible controls'. This maximum for $X_0 = i$ is denoted by $V(i)$. The `value function' $V(\cdot)$ then satisfies the dynamic programming equation
\begin{equation}
V(i) = \max_v[k(i,v) + \alpha\sum_jp(j|i,v)V(j)], \ i \in S. \label{DP}
\end{equation}
$V$ can be computed by, e.g., the value iteration algorithm
$$V_{n+1}(i) = \max_v[k(i,v) + \alpha\sum_jp(j|i, v)V_n(j)], \ i \in S,$$
beginning with any guess $V_0 \in \R^s$. In reinforcement learning, one seeks a data-driven analog of this, where the nonlinearity due to the `max' operator on the right causes problem because the conditional expectation is inside the nonlinearity.\footnote{except in special  circumstances that allow a `post-decision state' formulation, see \cite{Powell}.} This obstructs any kind of empirical conditional averaging. One way around is to define the Q-values as the expression in square brackets on the right in (\ref{DP}), i.e.,
$$Q(i,v) := k(i,v) +  \alpha\sum_jp(j|i,v)V(j), \ i \in  S, v \in A.$$
These satisfy their own dynamic programming equation
\begin{equation}
Q(i,v) = k(i,v) + \alpha\sum_jp(j| i,v)\max_wQ(j, w) \ \forall i,v. \label{Q-DP}
\end{equation}
In turn, this can be solved by the `Q-value iteration'
$$Q_{n+1}(i,v) = k(i,v) + \alpha\sum_jp(j|i,v)\max_wQ_n(j, w) \ \forall \ i,v,$$
where now the conditional expectation w.r.t.\ $p(\cdot | i,v)$ is outside the max. This facilitates a data driven learning (or stochastic approximation) version as follows. First, when the current state is $X_n$ and the control chosen is $U_n$,  one replaces the conditional expectation on the right hand side by an evaluation at the (real or simulated) next state $X_{n+1}$, i.e.,
$$Q_{n+1}(i,v) = k(i,v) + \alpha\max_wQ_n(X_{n+1}, w), \  i = X_n, v = U_n,$$
leaving $Q(j, w), j \neq i$ or $w \neq u$ unchanged. The scheme is then stabilized by making it incremental, i.e., by replacing the full move suggested by the right hand side by a convex combination of it with the previous iterate $Q_n(i, u)$, with a small weight $a(n) \in (0, 1)$ on the former. This leads to the classical Q-learning algorithm
\begin{eqnarray}
\lefteqn{Q_{n+1}(i,v) = (1 - a(n)I\{X_n = i, U_n = v\})Q_n(i,v) \ +} \nonumber \\
&& a(n)I\{X_n = i, U_n = v\}(k(i, v) +  \alpha\max_wQ_n(X_{n+1}, w)) \nonumber \\
&=&  Q_n(i,v) + a(n)I\{X_n = i, U_n = v\}\times    \nonumber \\
&&(k(i, v) + \alpha\max_wQ_n(X_{n+1}, w) - Q_n(i, v)).  \label{Qlearn}
\end{eqnarray}
Here $I\{ \cdots \}$ is the `indicator function' which is $1$ if `$\cdots$' holds and $0$ otherwise. With the usual Robbins-Monro conditions on the stepsizes $\{a(n)\}$,
\begin{equation}
\sum_na(n) = \infty, \ \sum_na(n)^2 < \infty, \label{RM}
\end{equation}
this becomes a stochastic approximation algorithm. It is \textit{asynchronous} because only the component corresponding to the current state-action pair is updated at each time. A variant is
\begin{eqnarray}
\lefteqn{Q_{n+1}(i,v) =  Q_n(i,v) + a(\mu(i, v, n))I\{X_n = i, U_n = v\} } \nonumber \\
&&\times \ (k(i, v) + \alpha\max_wQ_n(X_{n+1}, w) - Q_n(i, v)),  \label{Qlearn2}
\end{eqnarray}
where $\mu(i, v, n) = \sum_{m=0}^nI\{X_m = i, U_m = v\}$ is the `local clock' at $(i, v)$. Suppose that for some $\delta > 0$,
\begin{equation}
\liminf_{n\uparrow\infty}\frac{\mu(i,v,n)}{n} \geq \delta  \ \ \forall \ i, v, \ \mbox{a.s.} \label{frequent}
\end{equation}
This ensures `sufficient exploration' in a precise sense, i.e., all state-action pairs are sampled `comparably often'. Under some additional restrictions on $\{a(n)\}$ (see Chapter 7, \cite{Borkar}), (\ref{Qlearn2}) tracks the ordinary differential equation (ODE)
\begin{equation}
\dot{Q}(t) = F(Q(t)) - Q(t) \label{ode-basic}
\end{equation}
where $F(x) := [F_{i,v}(x), i \in S, v \in A],$ suitably vectorized, is the dynamic programming operator for Q-values given by
\begin{equation}
F_{i,v}(x) = k(i, v) + \alpha\sum_jp(j| i, v)\max_wx_{j,w} \label{DPop}
\end{equation}
for $x = [x_{i,v}, i \in S, v \in A]$, also suitably vectorized. It can be seen that $F$ is a max-norm contraction:
$$\|F(x) - F(y)\|_\infty \leq \alpha\|x - y\|_\infty,$$
and therefore has a unique fixed point $Q^*$ that satisfies (\ref{Q-DP}). This can be shown to be the globally asymptotically stable equilibrium of (\ref{ode-basic}), from which the a.s.\ convergence of $\{Q_n\}$ to $Q^*$ can be inferred (see \cite{Borkar}, pp.\ 129-130). A similar analysis also aplies to (\ref{Qlearn}) except that the limiting ODE
becomes
\begin{eqnarray*}
\dot{Q}(t) &=& \Lambda(t)(F(Q(t)) - Q(t)) \nonumber \\
&=& \widetilde{F}_t(Q(t)) - Q(t), \label{ode-t} 
\end{eqnarray*}
where:\begin{itemize}
\item  $\Lambda(t)$ for each $t$ is a diagonal matrix with positive entries on the diagonal that reflect the relative time scaling due to asynchrony (\cite{Borkar}, Chapter 7), which can be shown to  be bounded away from zero a.s.\ under (\ref{frequent}), and,

\item  $\widetilde{F}_t(x) := (I - \Lambda(t))x + \Lambda(t)F(x)$ which is an $\| \cdot \|_\infty$-contraction with contraction coefficient $(1 - \delta(1 - \alpha))$ and a unique common fixed point $Q^*$.
\end{itemize}
 This can be analyzed similarly to (\ref{ode-basic}), but we avoid these complications and stick to (\ref{ode-basic}), because they are not central to our main goals here.

\subsection{Prospect Theory}

Expected Utility Theory assumes that individuals behave rationally in order to maximize their expected utility. On the other hand, prospect theory  \cite{KT, Fox, Ruggeri, Wakker} aims to describe the actual, empirically validated behavior of people. Prospect theory replaces the utility function with a valuation map over gains and losses defined with respect to a reference point, which is the inflection point of the S-shaped curve in our case. The marginal impact of change in value diminishes with distance from the reference point. Similar to expected utility theory, concavity for gains contributes to risk aversion for gains. On the other hand, convexity for losses contributes to risk seeking behavior.\footnote{In economics, concave utility function $u(\cdot)$  implies risk aversion because for stochastic returns denoted by $X$, Jensen's inequality leads to $E[u(X)] \leq u(E[X])$, meaning `sure returns' $E[X]$ fetch greater utility than random or risky returns $X$. Analogous statement applies to convex $u(\cdot)$ and risk seeking behavior.}

\begin{figure}[h!]
\begin{center}
 \includegraphics[width=0.5\columnwidth]{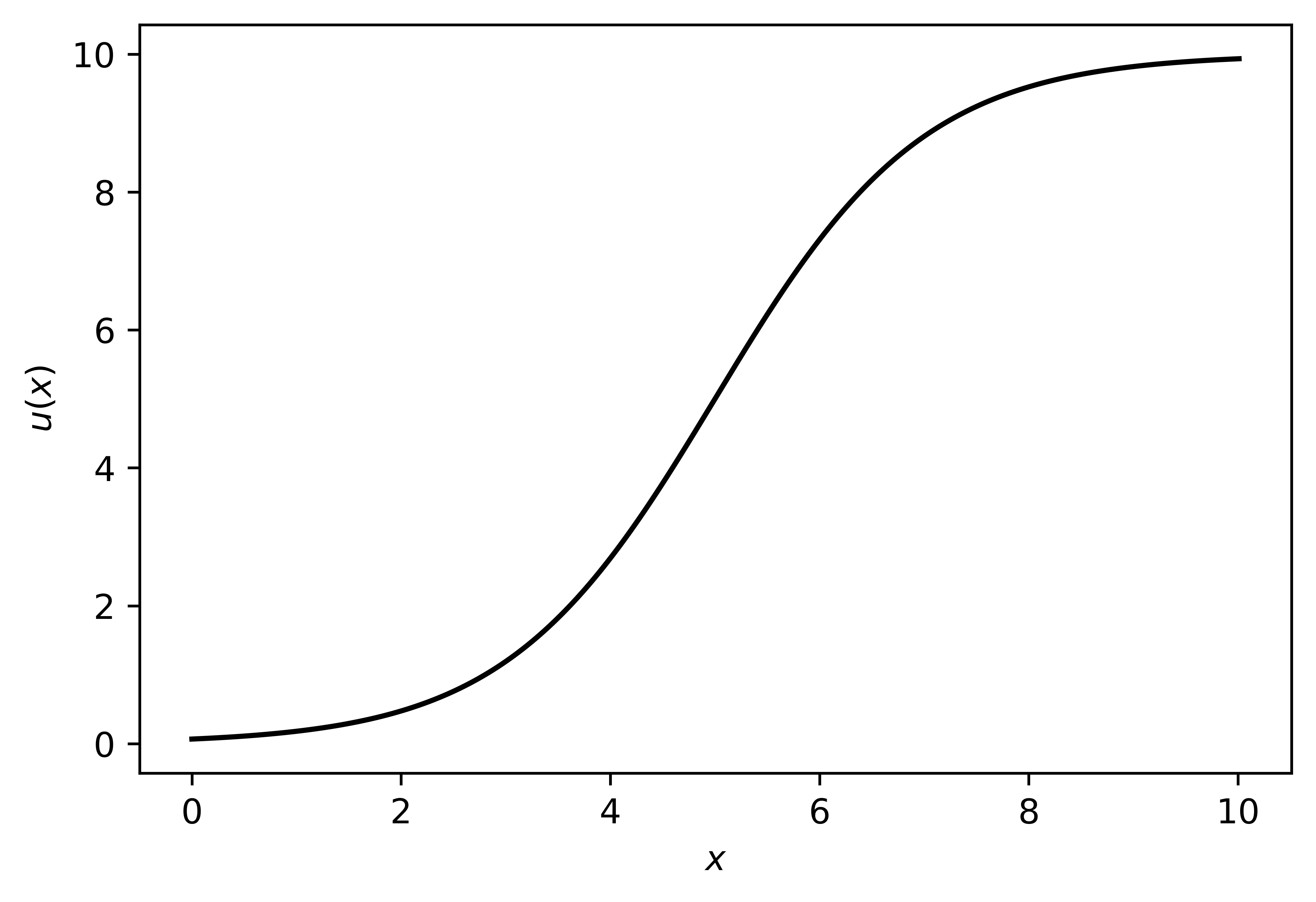}
 \caption{An example of the S-shaped map $u(\cdot)$ }
 \label{fig:u_example}
\end{center}
\end{figure}
We incorporate these ideas into the Q-learning scheme above by passing the estimated future returns through an S-shaped continuous and continuously differentiable map $u(\cdot)$ (see Figure \ref{fig:u_example}). What we take to be `estimated future returns' is, however, a non-unique choice. Some possibities are:
\begin{enumerate}
\item $\max_vQ(X_{n+1}, v)$. \textit{Prima facie}, this seems natural because this is precisely the term that depends on the next state, i.e., the `future'. This leads to the scheme
\begin{eqnarray}
\lefteqn{Q_{n+1}(i,v) =  Q_n(i,v) + a(\mu(i, v, n))I\{X_n = i, U_n = v\}}    \nonumber \\
&&\times \ (k(i, v) + \alpha u(\max_wQ_n(X_{n+1},w)) - Q_n(i, v)). \label{Qlearn3}
\end{eqnarray}

\item $k(i, v) + \alpha \max_wQ(X_{n+1}, w)$. This is the net estimate of future returns including the immediate reward.
This leads to
\begin{eqnarray}
\lefteqn{Q_{n+1}(i,v) =  Q_n(i,v) + a(\mu(i, v, n))I\{X_n = i, U_n = v\} \ \times}    \nonumber \\
&&(u(k(i, v) + \alpha\max_wQ_n(X_{n+1}, w)) - Q_n(i, v)). \label{Qlearn4}
\end{eqnarray}

\item $k(i, v) + \alpha \max_wQ(X_{n+1}, w) - Q_n(i,v)$. This is the estimate of the incremental future reward. This leads to
\begin{eqnarray}
\lefteqn{Q_{n+1}(i,v) =  Q_n(i,v) + a(\mu(i, v, n))I\{X_n = i, U_n = v\} \ \times} \nonumber \\
&& u(k(i,v) + \alpha\max_wQ_n(X_{n+1}, w) - Q_n(i, v)), \label{Qlearn2_alt}
\end{eqnarray}
This, however, goes against the spirit of the derivation of (\ref{Qlearn2}) in the preceding section because the update is no longer a convex combination of the previous iterate and a correction term. Nevertheless, it has been used in the literature \cite{ShenT}. We do not pursue this variant here.
\end{enumerate}

Our interest is in the qualitative analysis of the asymptotic behavior of such algorithms as reflected in the limiting o.d.e. Any of the above leads to a monotone dynamics and our analysis applies, though the actual locations of the equilibria may shift. This is confirmed by our numerical experiments. We mostly focus on the first model. 

In fact, we tweak even this model a little by adding noise. This is detailed in the next section.

\section{Prospect theoretic Q-learning}\label{sec:scheme_convergence}

\subsection{Modified Q-learning scheme}

Let $\{X_n\}, \{U_n\}$ be as above. We shall make the additional assumption that the graph of the Markov chain remains irreducible under all control choices at the nodes.   Consider the prospect theoretic scheme:
\begin{eqnarray}
\lefteqn{Q_{n+1}(i,v) = Q_n(i,v) + a(\mu(i,v,n))I\{X_n = i, U_n = v\}\Big(k(i,v) \ +} \nonumber \\
&& \alpha u(Q_n(X_{n+1}, U_{n+1}) - \xi_n(X_{n+1}, U_{n+1})) - Q_n(i,v)\Big) 
\label{QL}
\end{eqnarray}
where $\{\xi_n = [\xi_n(i,v)]\}$ is $\R^{sr}$-valued zero mean i.i.d.\ noise. Each $\xi_n(i,v)$ is distributed according to a continuously differentiable density $\varphi(\cdot)$ concentrated on a finite interval $[-c, c]$ for some $c \in [0, k_{min}]$. $U_{n+1}$ is chosen according to an epsilon-greedy policy, i.e., 
for a prescribed $\epsilon \in (0,1)$,
\begin{equation} \label{epsilon-greedy}
    U_{n+1}=\begin{cases} 
      w_{n+1}^* & \mathrm{w.p.} \ (1-\epsilon) \\
      w\neq w_{n+1}^* & \mathrm{w.p. } \ \frac{\epsilon}{r-1} \  \mathrm{ each}
   \end{cases}
\end{equation}
where $w_{n+1}^*=\mathrm{arg} \max_w (Q_n(X_{n+1}, w) - \xi_n(X_{n+1}, w))$, with ties broken by choosing a maximizer with equal probability. We ignore the latter possibility henceforth for sake of simplicity.

The `noise' $\{\xi_n\}$ can be justified as being caused by limited information, noisy measurements, etc. It serves a useful mathematical purpose  here, as we note later.

Define $K := \frac{k_{max}}{1 - \alpha}$ and let $u: [0, K+c] \mapsto [0, K]$ be continuously differentiable.

\begin{lemma}
\label{iteration-bounded}
If $Q_0 \in \SA:=[k_{min},K]^{sr}$, then $Q_n \in \SA \ \forall \ n \geq 0$.
\end{lemma}
\begin{proof}
Note that the Q-learning iteration can be written as:
\begin{align}\label{QL_convex}
Q_{n+1}(i,v) &= \Big(1-a(\mu(i,v,n))I\{X_n = i, U_n = v\}\Big)Q_n(i,v) \nonumber\\
&+ a(\mu(i,v,n))I\{X_n = i, U_n = v\}\Big(k(i,v) \ + \nonumber \\
& \alpha u(Q_n(X_{n+1}, U_{n+1}) - \xi_n(X_{n+1}, U_{n+1}))\Big). 
\end{align}
So $Q_{n+1}(i,v)$ is a convex combination of 
$$k(i,v) + \alpha u(Q_n(X_{n+1}, U_{n+1}) -\xi_n(X_{n+1}, U_{n+1}))$$ 
and $Q_n(i,v)$. Now, if $Q_n\in[k_{min},K]^{sr}$, then
\begin{align*}
k_{min}&\leq k(i,v)+\alpha u(Q_n(X_{n+1}, U_{n+1}) -\xi_n(X_{n+1}, U_{n+1})) \\
&\leq k_{max}+\alpha u(K+c) \nonumber\\
&=k_{max}+\alpha K \nonumber\\
&=K. \nonumber
\end{align*}
Hence $Q_n\in\SA\Rightarrow Q_{n+1}\in\SA$. The claim follows.
\end{proof}
We assume that the Q-learning iteration was initiated in the set $\SA$. As shown in Chapter 2 of ~\cite{Borkar}, since $u(\cdot)$ is Lipschitz continuous (it is continuously differentiable with bounded first derivative), and $\sup_n\|Q_n\|_\infty\leq K<\infty$, the iteration (\ref{QL}) almost surely tracks the asymptotic behavior of the o.d.e. :
\begin{eqnarray}
\lefteqn{\frac{d}{dt}q_t(i,v) = h_{i,v}(q_t)} \nonumber \\
&:=& F_{i,v}(q_t) - q_t(i,v)  \nonumber \\
&:=& k(i,v) + \nonumber \\
&&\alpha\int_{\R^{sr}} \bigg(\sum_jp(j|i,v)\Big((1-\epsilon)\max_{w'}\big(u(q_t(j,w') - y_{j,w})\big) \nonumber \\ 
&+& \frac{\epsilon}{r-1}\sum_{w\neq w^*_{q_t,y,j}}\big(u(q_t(j,w) - y_{j,w})\big) \Big) \bigg) \ \times \nonumber \\
&&\prod_{j,w} \varphi(y_{j,w})dy_{j,w} - q_t(i,v).  \ 
\label{ode}
\end{eqnarray}
where $w^*_{q_t,y,j}=\mathrm{arg} \max_w (q_t(j, w) - y_{j,w})$.
The above implicitly defines the maps  $h, F: \R^d \mapsto \R^d$, in fact, $h(x) = F(x) - x$. The integral is a convolution w.r.t.\ a continuously differentiable function, which makes it continuously differentiable.

\begin{lemma}\label{ode-bounded}
When initiated in the set $\SA:=[k_{min},K]^{sr}$, o.d.e.\ (\ref{ode}) stays in the set $\SA$.
\end{lemma}
\begin{proof}
We can apply the following inequality on the derivative of $q_t(i,v)$:
\[
k_{min}-q_t(i,v)\leq\frac{d}{dt}q_t(i,v)\leq k_{max}+\alpha u(K+c)-q_t(i,v)
\]
Its discretization can be written as:
\[
a_nk_{min}+(1-a_n)q_n(i,v)\leq q_{n+1}(i,v) \leq a_nK+(1-a_n)q_n(i,v)
\]
This implies that if $q_n(i,v)\in[k_{min},K]$, then $q_{n+1}(i,v)\in[k_{min},K]$. So, if initiated in the set $\SA := [k_{min}, K]^{sr}$, $q_n$ (and its limit, the o.d.e.) stays in the set $\SA$.
\end{proof}

The Jacobian matrix of $h$ (resp., $F$) at $q$ is $J(q) - I$ (resp., $J(q)$), where $I$ is the $sr\times sr$ identity matrix and $J(q)$ for $q \in \R^{sr}$ is the $sr\times sr$ matrix whose $((i,v), (j,w))$\textsuperscript{th} element is
\begin{eqnarray}
\lefteqn{J(q)_{(i,v),(j,w)}= p(j|i, v)\alpha \int\Bigg\lbrack\Big((1-\epsilon)u'(q(j,w) - y_{j,w}) \ \times } \nonumber \\
&&I_{\{q(j,w) - y_{j,w} > q(j,w') - y_{j,w'} \ \forall \ w' \neq w\}} + \frac{\epsilon}{r-1}u'(q(j,w) - y_{j,w})\  \nonumber \\
&&\times \ \big(1-I_{\{q(j,w) - y_{j,w} > q(j,w') - y_{j,w'} \ \forall \ w' \neq w\}}\big)\Big) \prod_{w} \varphi(y_{j,w})dy_{j,w}\Bigg\rbrack. \nonumber \\
&& \ 
\label{Jacobian}
\end{eqnarray}
The continuous differentiability of $h$ was facilitated by the presence of the noise $\{\xi_n\}$, which allows us to exploit the theory of monotone dynamical systems that is available when it holds. For $h$ that are only Lipschitz, such theory appears to be lacking.

\subsection{Monotonicity and its consequences}\label{sec:monotone}

We use the following notion of cooperative o.d.e.\ from \cite{Smith}:
\begin{definition} (Cooperative o.d.e.)
An o.d.e.\ of the form $\dot{x}=h(x(t))$ is a cooperative o.d.e.\ if 
\[
\frac{\partial{h_i}}{\partial{x_j}}\geq0, ~ j\neq i.
\]
and the Jacobian matrix for $h$ is irreducible.
\end{definition}

\begin{lemma}\label{cooperative}
When the controlled Markov chain is irreducible, $J(q)$ (the Jacobian of F) is a non-negative irreducible matrix and o.d.e.\ (\ref{ode}) is a cooperative o.d.e.
\end{lemma}
\begin{proof}
Since $u' > 0$, it follows that $J(q)$ is a non-negative matrix. $J(q)$ can be written as the product of the following two matrices: a non-negative $sr\times sr$ matrix $P$ where $P_{(i,v),(j,w)}=p(j|i,v)$ and a positive diagonal matrix $J_1(q)$, where $J_1(q)_{(j,w),(j,w)}$ is $\alpha$ times the integral in equation (\ref{Jacobian})). Since the Markov chain is irreducible, the matrix $P$ is irreducible and hence, the matrix $J(q)$ will be irreducible. Since the off-diagonal terms of $J(q)-I$ are non-negative and the Jacobian is irreducible, (\ref{ode}) is a cooperative o.d.e. 
\end{proof}

\begin{corollary} The dynamical system described by (\ref{ode}) is \textit{monotone} in the sense that if $q(\cdot), q'(\cdot)$ are two solutions with $q(0) \geq q'(0)$ componentwise, then $q(t) \geq q'(t) $ componentwise $\forall \ t \geq 0$.\end{corollary}

This follows from the results of \cite{Hirsch}.  The next theorem follows as a consequence of Theorem 2.1 from \cite{Hirsch1}
\begin{theorem}\label{BigThm}
For initial conditions in an open dense set, the solutions of (\ref{ode}) converge to an equilibirium.
\end{theorem}
The same must then be true for the iterates of the discrete map $\Phi : \SA \mapsto \SA$ which maps $q_0$ to $q_1$. Since (\ref{ode}) is cooperative, this map is monotone. It is also `order compact' in the sense of \cite{Hirsch2}, section 5.1, because it maps an order interval, i.e., a set of the form $[y,z]^* := \{ x : y \leq x \leq z$ componentwise$\}$ to a bounded set. We define $[a, b)^*$ etc.\ analogously. Using part (b) of Theorem 5.6 from \cite{Hirsch2}, the following holds:
\begin{theorem}\label{max_min_equi} There exist maximal and minimal equilibria $q^*, q_*$ resp.\ of (\ref{ode}), such that any other equilibrium $\hat{q}$ of (\ref{ode}) satisfies $q_* \leq \hat{q} \leq q^*$  componentwise.
\end{theorem}
Since the dynamics preserves order, it follows that for $q_t, t \geq 0,$ satisfying (\ref{ode}), $q_0 \geq q^* \Longrightarrow q_t \geq q^*$ and likewise, $q_0 \leq q_* \Longrightarrow q_t \leq q_*$. If $q^* > q_*$, since both are fixed points (i.e., equilibria) for the dynamics, $q_* \leq q_0 \leq q^* \Longrightarrow q_* \leq q_t \leq q^* \ \forall \ t \geq 0$ by monotonicity. Since the map $q_0 \mapsto q_1$ is continuous, the following theorem and its corollary hold:
\begin{theorem}\label{trichotomy}
(Order Interval Trichotomy, Theorem 5.1 from \cite{Hirsch2}) At least one of the following holds:

\begin{enumerate}
\item $\exists$ a third equilibrium $\hat{q}$ such that  $q_* < \hat{q} < q^*$,

\item $\exists$ a trajectory $q_t$ of (\ref{ode}) such that $q_t\uparrow q^*$ as $t\uparrow\infty$ and $q_t\downarrow q_*$ as $t\downarrow-\infty$,

\item  $\exists$ a trajectory $q_t$ of (\ref{ode}) such that $q_t\downarrow q_*$ as $t\uparrow\infty$ and $q_t\uparrow q^*$ as $t\downarrow-\infty$.
\end{enumerate}\end{theorem}
\begin{corollary}\label{trichotomy_stable}
(Corollary 5.2 from \cite{Hirsch2}) If both  $q_*$ and $q^*$ are stable, then there is at least one more equilibrium $\hat{q}$ such that $q_* < \hat{q} < q^*$.
\end{corollary}

\section{Equilibrium Points}\label{eq.points}
\subsection{Regions with stable equilibria}\label{sec:stable_regions}

As our discrete map is the time-$1$ map of a differential equation with smooth right hand side, the stability of its equilibria, which are the same as equilibria of the differential equation, can be analyzed by looking at the  linearization of $h(\cdot) := [h_{i,v}(\cdot)]_{i \in S, v \in A}$ at the equilibrium, i.e., the eigenvalues of the Jacobian matrix $J(q)-I$ evaluated at the equilibrium.
We have the following interesting observation from the foregoing.

\begin{theorem} If $q^*$, resp.\ $q_*$, is hyperbolic, then it is stable. \end{theorem}

\begin{proof} If (say) $q^*$ is hyperbolic, it is isolated by the inverse function theorem and has well defined stable and unstable manifolds in an open neighborhood, corresponding to the eigenvalues of $J(q^*) - I$ in the left, resp.\ right half of the complex plane. Furthermore, if the dimension of the unstable manifold is $\geq 1$ (i.e., there is at least one unstable eigenvalue), then for all initial conditions in an open neighborhood $O$ of $q^*$, the trajectories that are not initiated exactly on the stable manifold eventually move away from $q^*$. Since the dimension of the stable manifold is strictly less than $s$, this is so for initial conditions in an open dense subset $O'$ of $O$. We may further exclude from $O'$ its intersection with the closed nowhere dense set where the convergence claim of Theorem \ref{BigThm} can fail and denote the resultant open dense  subset of $O$ as $O''$. Denote the intersection of $O''$ with the open cone $C^* := \{q : q = q^* + y, y > 0$ componentwise$\}$, as $\hat{O}$, which will be an open set. Then $q(0) \in \hat{O}$ implies that $q(t)$ eventually moves away from $q^*$. By monotonicity,  $q(0) \geq q^*$  componentwise $\Longrightarrow q(t) \geq q^* \ \forall \ t \geq 0$ componentwise by monotonicity. Since $q^*$ is the maximal equilibrium, if $q(0) \in \hat{O}$, it must hold that $q(t) \to q^*$, a contradiction to the previous claim. Hence $q^*$ must be stable. The claim for $q_*$ is proved similarly. \end{proof}

 The following theorem gives bounds on the eigenvalues of $J(q)-I$:

\begin{theorem}\label{perron}
(Perron-Frobenius Theorem \cite{nonneg}) Let A be a square non-negative irreducible matrix. Then
\begin{enumerate}
    \item $A$ has a real positive eigenvalue $\lambda_A$ which is greater than or equal to the absolute value of any other eigenvalue of $A$.
    \item $r\leq\lambda_A\leq R$ where $r=\min_ir_i$ and $R=\max_ir_i$, where $r_i$ denotes the sum of the elements of row $i$ of $A$.
\end{enumerate}
\end{theorem}

Let the sum of the $(i,v)^{\textrm{th}}$ row of $J(q)$ be $\Gamma(q)_{i,v}$. Let $\Gamma(q)^*=\max_{i,v}\Gamma(q)_{i,v}$ and  $\Gamma(q)_*=\min_{i,v}\Gamma(q)_{i,v}$. Since $J(q)$ is a non-negative irreducible matrix, there exists a real positive eigenvalue $\lambda^*$ (dominant eigenvalue) for $J(q)$ such that any other eigenvalue $\lambda$ of $J(q)$ has its absolute value (and hence its real part) smaller than or equal to $\lambda^*$, by Theorem \ref{perron}. We also know that $\Gamma(q)_*\leq \lambda^* \leq \Gamma(q)^*$. For any eigenvalue $\lambda$ of $J(q)$, $\lambda-1$ is an eigenvalue of the Jacobian $J(q)-I$. So, the real part of all eigenvalues of $J(q)-1$ are less than $\lambda^*-1$.

Since $u(\cdot)$ is an S-shaped function, we know that $u'(x)$ is typically small for sufficiently low or high values of $x$. Thus $u'(x) < 1 < \frac{1}{\alpha}$ for sufficiently low or high values of $x$ and can exceed $\frac{1}{\alpha}$ in the mid-range. If $u'(x)<\frac{1}{\alpha} \ \forall x\in[0,K+c]$, then we can use the results from \cite{ShenS} which show that there will exist only one equilibrium point in the set and it will be stable. In fact this follows easily from the contraction mapping theorem. We consider here the case where $u'(x)$ exceeds $\frac{1}{\alpha}$ in the middle region. Define points $a,b, b > a,$ in $[0,K]$ as the largest and smallest points, respectively, in $[0,K]$ such that $u'(x)<\frac{1}{\alpha} \ \forall x\in [0,a)\cup(b,K+c]$.

\begin{theorem}
\label{theorem:stable}
There is at most one equilibrium point for (\ref{ode}) in the set $(b+c,K]^{sr}$. If such an equilibrium point exists, it will be a stable equilibrium and the maximal equilibrium point. Similarly, there is at most one equilibrium point for (\ref{ode}) in the set $[k_{min},a-c)^{sr}$ and if such an equilibrium point exists, it will be a stable equilibrium and the minimal equilibrium point.
\end{theorem}
\begin{proof}
Let $q_1$ be any point in the set $(b+c,K]^{sr}$. Then $[q_1-c,q_1+c]^*\subseteq(b,K+c]$ componentwise. Since $u'(\cdot)<1/\alpha$ in this region, $\Gamma(q_1)_{i,v}<1, \forall i,v$ and hence, $\lambda^*<1$. Hence real parts of all eigenvalues of the Jacobian $J(q_1)-I$ are negative. So any equilibrium point lying in this region will be hyperbolic and stable. 
Now suppose that there are two equilibria $q_1, q_2$ in the aforementioned set. The two points can be ordered or unordered. First we consider the case where they are ordered and $q_1<q_2$. By Corollary \ref{trichotomy_stable}, there exists another equilibrium point $q_3$, such that $q_1<q_3<q_2$ . Then $q_3$ will also be a stable equilibrium and hence there will be more stable equilibrium points between $q_1, q_3$, and between $q_3, q_2$. Repeated application of this argument implies that we will have a continuum of non-isolated equilibria. But real part of all eigenvalues of the Jacobian $J(q)-I$ are negative in this region, implying that all equilibria are isolated. This gives us a contradiction. Hence there cannot be two ordered equilibria in the region.

Now consider the other case in which there are two unordered equilibria $q_1, q_2$ in the region $(b+c,K]^{sr}$. We know that there exists $q^*$ such that all equilibrium points $q$ satisfy $q\leq q^*$ (Theorem \ref{max_min_equi}). Since no ordering exists between $q_1$ and $q_2$, they can't be equal to $q^*$. So, $q_1<q^*$ where both $q_1$ and $q^*$ lie in this region. But we have shown earlier that there cannot exist ordered equilibria in the region. So, at most one equilibrium point can exist in this region and that will be the maximal equilibrium point.
Analogous statement  for the set $[k_{min},a-c)^{sr}$ is proved similarly.
\end{proof}
Based on Theorem \ref{theorem:stable}, we subsequently refer to the sets $[k_{min},a-c)^{sr}$ and $(b+c,K]^{sr}$ as the lower and upper stable regions, respectively. 

\subsection{Additional results on stability of equilibria}

Let points $d,e$ in $[0,K]$ be the smallest and the largest points in $[0,K]$ such that $u'(x)>\frac{1}{\alpha} \ \forall \ x, d < x < e$. In most cases, $d$ and $e$ will be same as $a$ and $b$ respectively, but we have defined them separately to take care of cases where $u'(x)=\frac{1}{\alpha}$ in the intervals $x\in[a,d]$ and $x\in[e,b]$. 
\begin{theorem}
Any equilibrium point in the region $(d+c,e-c)^{sr}$ is an unstable equilibrium point.
\end{theorem}
\begin{proof}
Let $q_1$ be any point in the set $(d+c,e-c)^{sr}$. Then, $[q_1-c,q_1+c]^*\subseteq(d,e)$ componentwise. Since $u'(\cdot)>1/\alpha$ in this region, $\Gamma(q_1)_{i,v}>1, \forall i,v$ and hence, $\lambda^*>1$. Thus at least one eigenvalue of $J(q) - I$ has a positive real part. Therefore any equilibrium point in this region will be unstable. 
\end{proof}

Let the maximum value of $u'(x)$ be attained at $m_1$.  
\begin{theorem}
If all equilibrium points are hyperbolic and $u(x)$ is convex, respectively, concave in the regions $x<m_1$ and $x>m_1$, respectively, then there can exist at most one stable equilibrium point in the region $[k_{min},m_1-c)^{sr}$. Similarly in the region $(m_1+c,K]^{sr}$, there can exist at most one stable equilibrium. If these exist, then they will be the minimal and maximal equilibrium points, respectively. 
\end{theorem}
\begin{proof}
Suppose there exist two stable equilibrium points $q_1,q_2$ in the region $(m_1+c,K]^{sr}$. Again, there are two possibilities: they can be ordered or unordered. Let us first consider the case where they are ordered, say $q_1 < q_2$. Since we have taken $u(\cdot)$ to be concave in the region, for any point $q_3\in[q_1,q_2]^*$, $u'(q_1(i,v) - y)\geq u'(q_3(i,v) - y)\geq u'(q_2(i,v) - y)$ $\forall \ y \in [-c, c]$. Hence $\Gamma(q_1)\geq \Gamma(q_3) \geq \Gamma(q_2)$ componentwise. Let $\lambda^*_{q_1}, \lambda^*_{q_2}$ and $\lambda^*_{q_3}$ be the dominant eigenvalues of $J(q_1), J(q_2)$ and $J(q_3)$ respectively. Then $\Gamma(q_1)\geq \Gamma(q_3) \geq \Gamma(q_2)$ implies that $\lambda^*_{q_1}\geq \lambda^*_{q_3}\geq\lambda^*_{q_2}$ (Theorem A.9 from \cite{nonneg}). Since $q_1$ and $q_2$ are stable equilibria, $\lambda^*_{q_1}$ and $\lambda^*_{q_2}$ are less than 1, and hence $\lambda^*_{q_3}<1$. So, $q_3$ will also be a stable equilibrium. As shown in the proof of Theorem \ref{theorem:stable}, this gives us a continuum of non-isolated equilibria, which contradicts our assumption that all equilibria are hyperbolic.

If there exist two unordered equilibrium points in the region, then there will exist another equilibrium point $q_3$ which will be the maximal equilibrium and will be stable, as we have assumed that all equilibria are hyperbolic. Applying the first part of this proof to $q_1$ and $q_3$ gives us a contradiction. Hence there can be at most one stable equilibrium point in the region $(m_1+c,K]^{sr}$. The same is true for the set $[k_{min},m_1-c)^{sr}$.
\end{proof}
This theorem can also be applied where the valuation map is a traditional utility function which is either convex or concave in the whole domain. In our case, however, there can exist many other stable equilibrium points with some components below and some above $m_1$.

Let $b_1$ be a point such that $k_{min}+\alpha u(b_1-c)\geq b_1$. Then $k_{min}+\alpha u(x-c)\geq b_1 \  \forall x\geq b_1$. In the convex combination form of the iteration (\ref{QL_convex}), note that if $Q_n\geq b_1$, then
\begin{align*}
    & k(i,v) + \alpha u(Q_n(X_{n+1}, U_{n+1}) -\xi_n(X_{n+1}, U_{n+1})) \\
    &\;\;\geq  k_{min}+\alpha u(b_1-c) \geq b_1 
\end{align*}
and hence $Q_{n+1}\geq b_1$. Similar to Lemma \ref{iteration-bounded}, we note that if the iteration (\ref{QL}) is initiated in the set $[b_1,K]^{sr}$, then it will stay in this set. In the following theorem, we use this to provide a sufficient condition for a stable equilibrium point of (\ref{ode}) to exist in the upper stable region. For simplicity, define $u_1(x):=k_{min}+\alpha u(x-c)$.
\begin{theorem}
\label{theorem:u_1suff}
If $u_1(b+c)\geq b+c$, then there exists a stable maximal equilibrium point of (\ref{ode}) in the region $[b+c,K]^{sr}$ and when the iteration (\ref{QL}) is initiated in this set, it converges to this equilibrium point.
\end{theorem}
\begin{proof}
In the region $(b+c,K]$, $u_1'(x)<1$ and hence $u_1$ and $y=x$ can intersect at most once in the region. Such an intersection point would exist if $u_1(b+c)>b+c$. If $u_1(b+c)=b+c$, then that point will be $b+c$ and their will be no other intersection. Let this intersection point be called $g$. 

As stated before, when $u_1(b+c)\geq b+c$, then the iteration (\ref{QL}) would stay in the set $[b+c,K]^{sr}$ when initiated in it. Similarly, if initiated in the set $[g,K]^{sr}$, the iteration stays in that set. Hence there exists an equilibrium point of (\ref{ode}) in this set. As proved in Theorem \ref{theorem:stable}, there can be at most one equilibrium point in the set $[b+c,K]^{sr}$. Hence, if $u_1(b+c)\geq b+c$ the equilibrium point in the set $[g,K]^{sr}$ will be the stable maximal equilibrium point and whenever iteration (\ref{QL}) is initiated in the region $[b+c,K]^{sr}$, it will converge to this equilibrium point.
\end{proof}

Similar to $u_1(\cdot)$, define $u_2(x):=k_{max}+\alpha u(x+c)$. We have the following theorem relating $u_2(\cdot)$ and the lower stable region similar to Theorem \ref{theorem:u_1suff}:
\begin{theorem}
\label{theorem:u_2suff}
If $u_2(a-c)\leq a-c$, then there exists a stable minimal equilibrium point of (\ref{ode}) in the region $[k_{min},a-c]^{sr}$ and when the iteration (\ref{QL}) is initiated in this set, it converges to this equilibrium point.
\end{theorem}

\begin{theorem}
If $u_1(a+c)> a+c$, then there exists only one equilibrium point of (\ref{ode}) in the set $[k_{min},K]^{sr}$ and it will lie in the region $(b+c,K]^{sr}$.
\end{theorem}
\begin{proof}
We know that $u_1'(x)=\alpha u'(x-c)\geq1, \forall x\in[a+c,b+c]$ (by the definition of $a,b$) and hence $u_1(a+c)> a+c$ implies that $u_1(b+c)>b+c$. From the proof of Theorem \ref{theorem:u_1suff}, we know that the curve $u_1(x)$ and $y=x$ intersect only once in the region $(b+c,K]$. Let this point of intersection be $g$. 
For all $x\in[a+c,b+c]$, $u_1'(x)\geq1$ and hence, $$u_1(x)-u_1(a+c)\geq x-(a+c) \implies u_1(x)\geq u_1(a+c)+x-(a+c)>x.$$ For all $x\in[k_{min},a+c]$, $u_1'(x)\leq1$ and hence, $$u_1(a+c)-u_1(x)\leq (a+c)-x \implies u_1(x)\geq u_1(a+c)-(a+c)+x>x.$$
So $u_1(x)$ lies strictly above $y=x$ in the region $[k_{min},b+c]$ and hence $g$ will be the only intersection of these two curves in the region $[k_{min},K]$. Consider any $q\notin[g,K]^{sr}$. Then its minimum component is in $[k_{min},g)$. Let this minimum component be $q_{i',v'}$ (i.e., $(i',v')=\argmin_{(i,v)} q_{i,v}$). Since $u_1(q_{i',v'})>q_{i',v'}$ and $q_{i,v}\geq q_{i',v'}, \forall (i,v)$, 
\begin{align*}
    h_{i',v'}(q)&\geq k_{i',v'}+\alpha u(q_{i',v'}-c)-q_{i',v} \\
    &\geq k_{min}+\alpha u(q_{i',v'}-c)-q_{i',v'}=u_1(q_{i',v'})-q_{i',v'} \\
    &>0.
\end{align*}
So there can be no equilibrium point of (\ref{ode}) which does not belong to $[g,K]^{sr}$. By Theorem \ref{theorem:u_1suff}, we know that there exists one stable equilibrium in the region $[g,K]^{sr}$. This will be the only equilibrium point and will be both the maximal and minimal equilibrium point of (\ref{ode}).
\end{proof}

\section{Numerical Experiments}\label{sec:simu}
We numerically simulated our modified Q-learning scheme and the corresponding o.d.e.\ to verify our results and gain additional insights. We used the shifted and scaled version of sigmoid logistic function:
\[
u(x)=\frac{L}{1+e^{-\gamma(x-x_0)}}.
\]
By modifying value of $L,\gamma$ and $x_0$, we explored the behavior of the Q-learning scheme with varying maximum value, steepness and midpoint of the curve, respectively. Both very small and large state and action spaces were explored with values of both $s$ and $r$ ranging from $2$ to $100$. To satisfy (\ref{RM}), we chose $a(\mu(i,v,n))$ as:
\[
a(\mu(i,v,n))=\frac{1}{\left\lceil\frac{\mu(i,v,n)}{100}\right\rceil}
\]
where $\lceil\cdot\rceil$ is the ceiling function and the scaling by $100$ was chosen for faster convergence. The rewards $k$ were generated randomly in a given range set by fixing $k_{min}$ and $k_{max}$. The transition matrix was generated randomly. We tried different values of the discount factor $\alpha$ ranging from $0.01$ to $0.99$. We used the raised cosine distribution for noise as we required the noise distribution to be continuously differentiable in a finite support. Noise was in general kept much smaller than the rewards $(c\approx0.01)$. Finally, we used $\epsilon=0.05$ for the $\epsilon$-greedy scheme. 

\begin{figure}[h!]
\centering
\begin{subfigure}{.45\textwidth}
 \includegraphics[width=\textwidth]{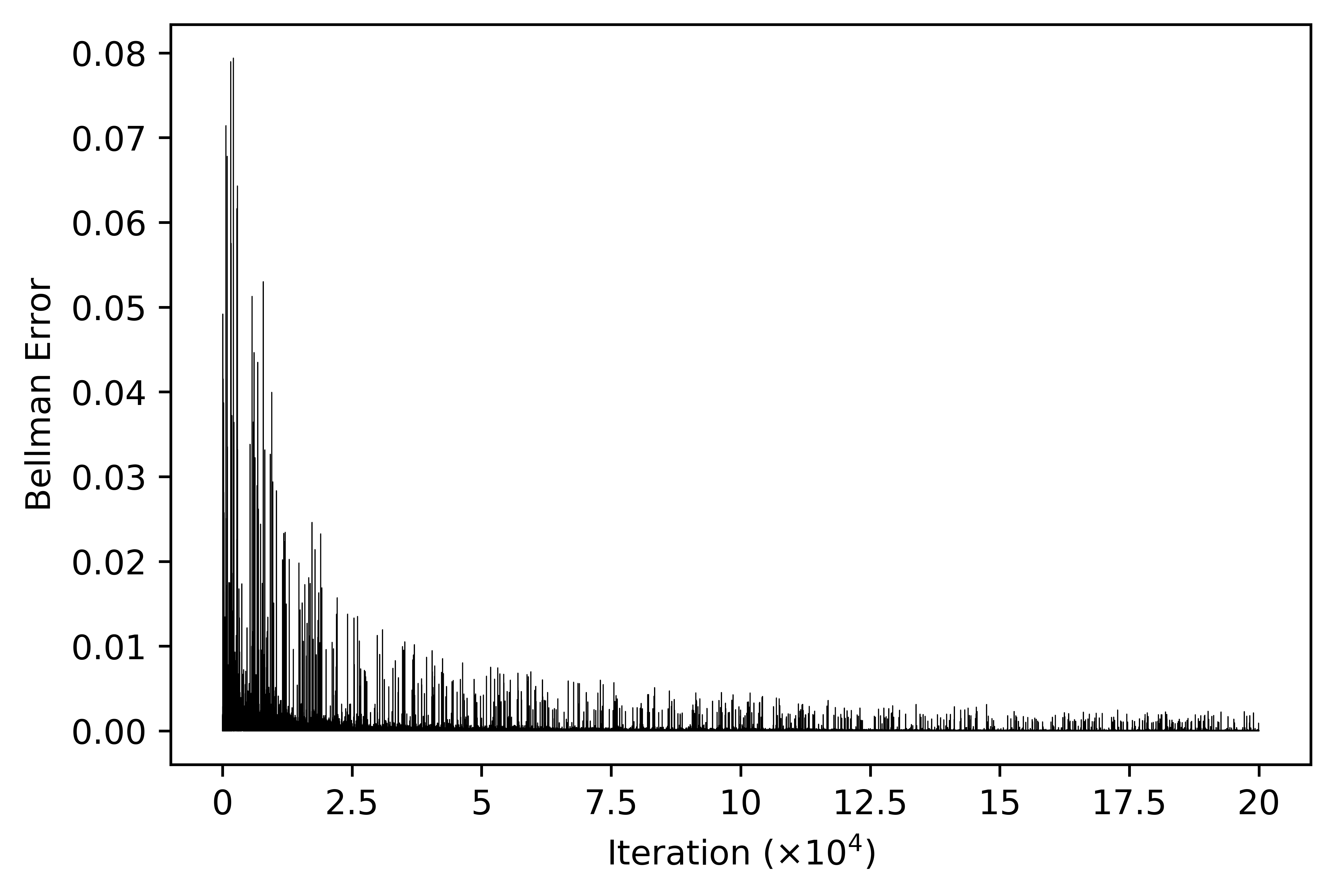}  
 \caption{}
 \label{fig:Bellman_direct}
\end{subfigure}
\begin{subfigure}{.45\textwidth}
 \includegraphics[width=\textwidth]{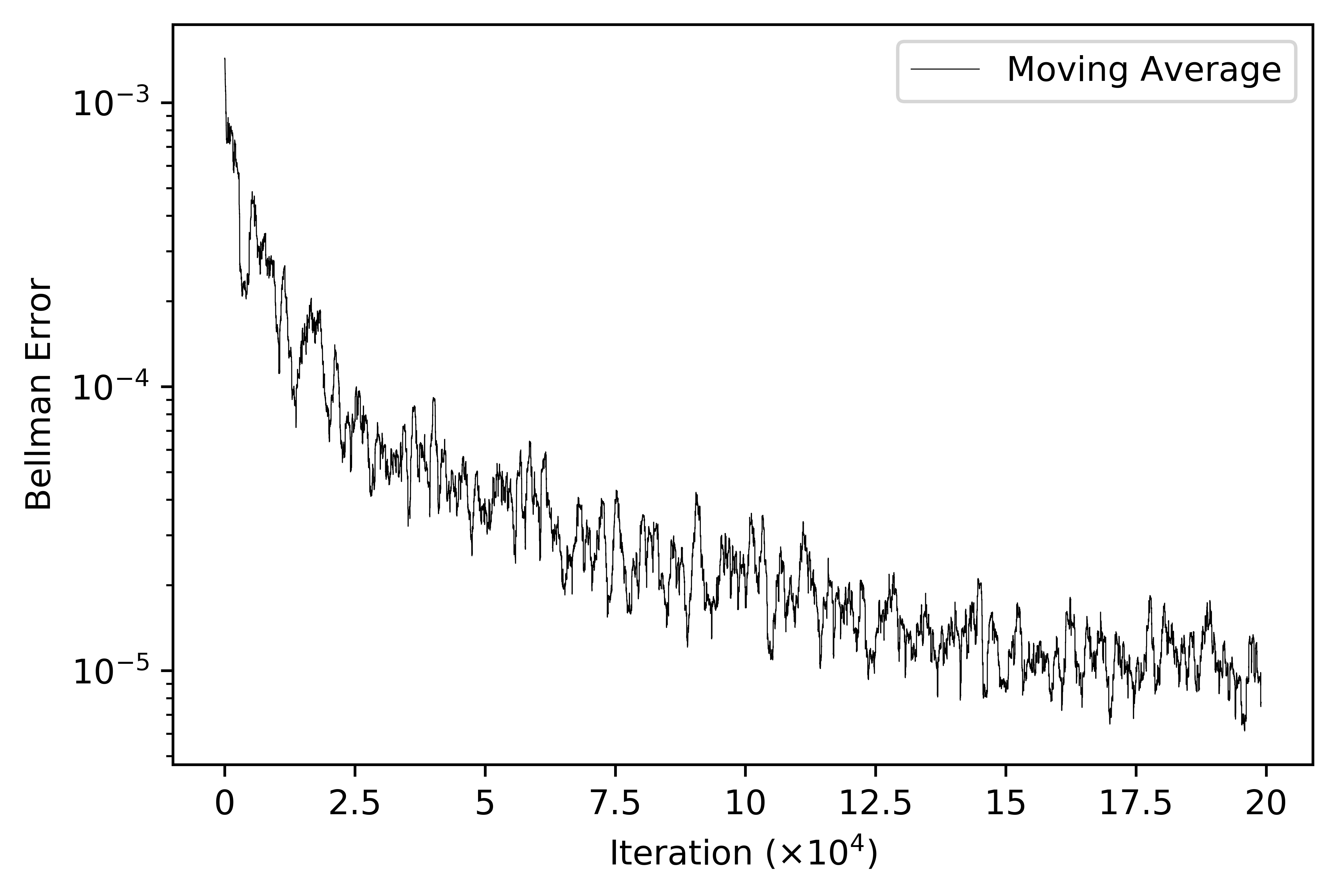}  
 \caption{}
 \label{fig:Bellman_moving_avg}
\end{subfigure}
\caption{Convergence plot of modified Q-learning scheme: (\protect\subref{fig:Bellman_direct}) shows the Bellman error for modified Q-learning scheme, while (\protect\subref{fig:Bellman_moving_avg}) shows the moving average of the same over 1000 iterations. ($\alpha=0.5, k_{min}=2, k_{max}=5,s=r=20,\epsilon=0.05$ and iteration initiated in $[K-1,K-0.5]^{sr}$)}
\label{fig:Bellman}
\end{figure}

We varied different parameters and studied the Q-learning iteration, the o.d.e.\ and their equilibrium points. These are some of our observations:
\begin{itemize}
    \item As long as the conditions mentioned in the Section \ref{sec:scheme_convergence} are met, the Q-learning iteration and the o.d.e.\ converged to an equilibrium point and to the same point when initiated at the same point. In Figure \ref{fig:Bellman}, we plot a representative convergence plot of  error vs.\ iterations where error for iteration $n$ is defined as: $|Q_{n+1}(X_n,U_n)-Q_{n}(X_n,U_n)|$. 
    \item  Choice of $\{a(n)\}$ and $\epsilon$ have an impact on the rate of convergence, but do not observably affect the equilibrium points.
    \item  As expected, when either $\alpha$ is too small or the function $u(\cdot)$ rises very gradually (i.e., $u'(x)<\frac{1}{\alpha}$ in the whole region), then there exists only one equilibrium point (see Figure \ref{fig:u}(\subref{fig:low_max_derivative})).  Even when $u'(x)$ exceeds $\frac{1}{\alpha}$ in the middle region, we frequently observe only one equilibrium point in $\SA$ when the maximum value of the first derivative of $u(\cdot)$ is not very high.

\begin{figure}[h!]
\centering
\begin{subfigure}{.43\textwidth}
 \includegraphics[width=\textwidth]{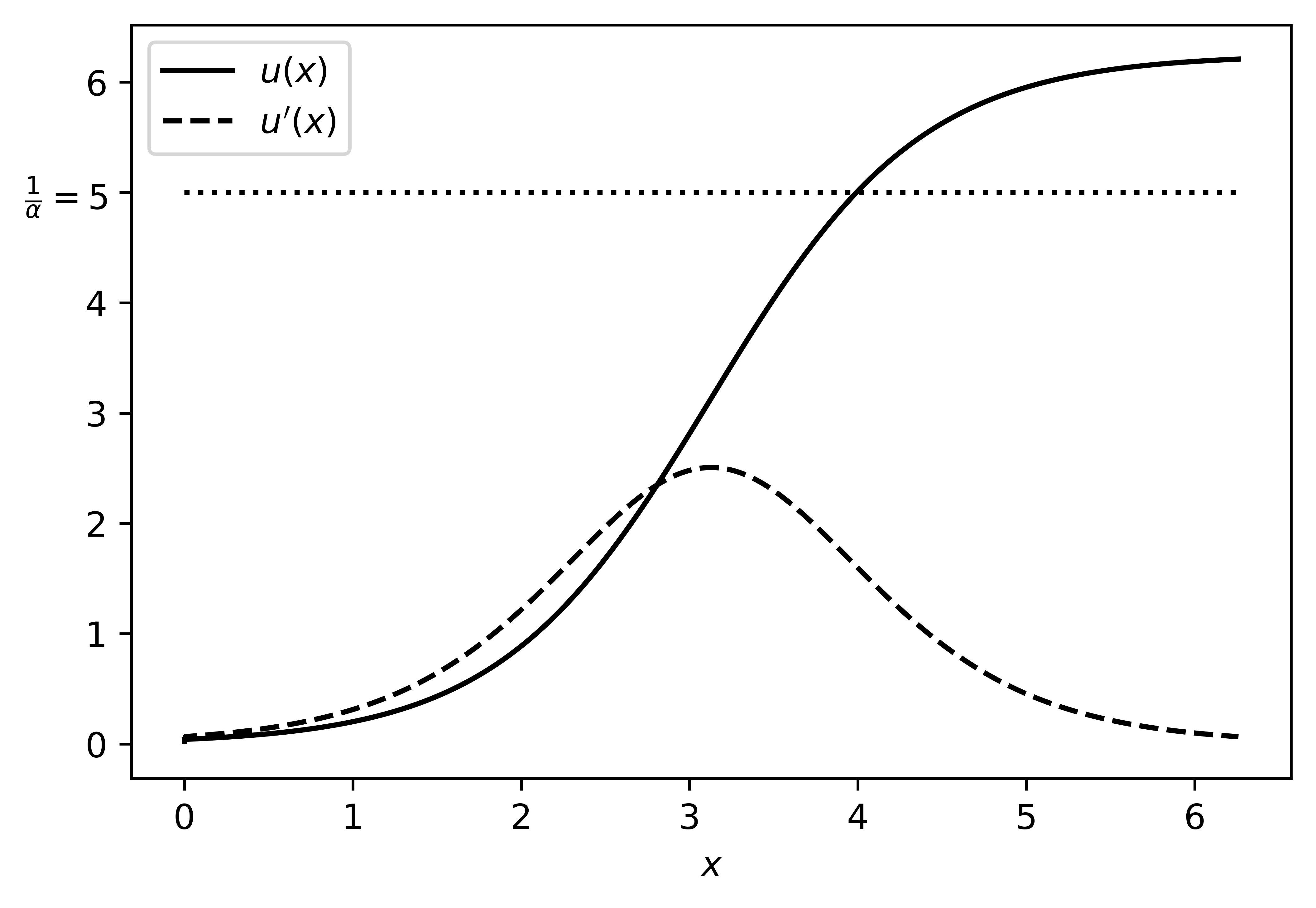}  
 \caption{}
 \label{fig:low_max_derivative}
\end{subfigure}
\begin{subfigure}{.45\textwidth}
 \includegraphics[width=\textwidth]{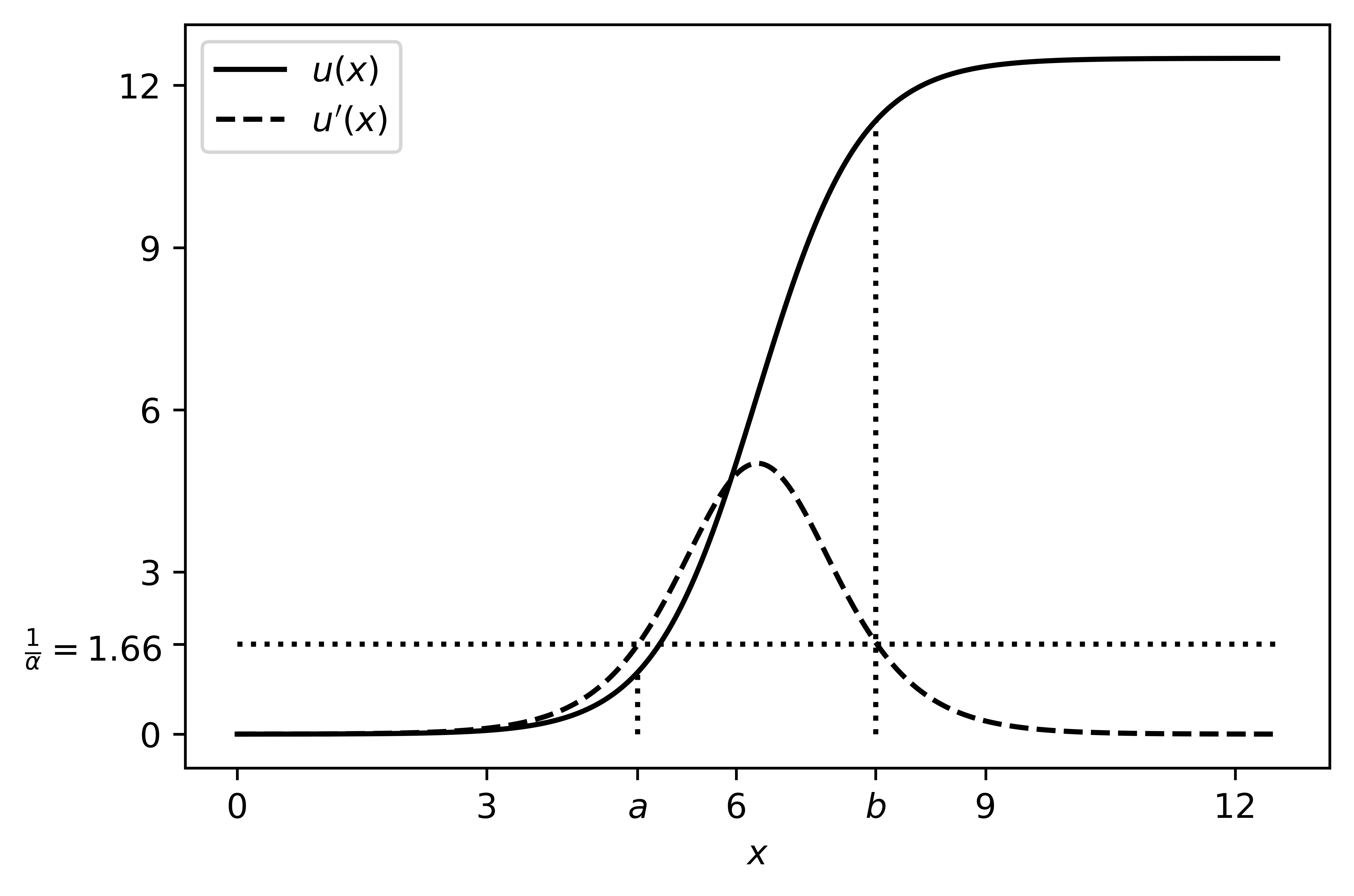}  
 \caption{}
 \label{fig:high_max_derivative}
\end{subfigure}
\caption{$u(\cdot)$ and $u'(\cdot)$ for two different cases: (\protect\subref{fig:low_max_derivative}) shows the case where $u(\cdot)$ rises gradually and $\max_xu'(x)<1/\alpha, (\alpha=0.2)$. Only one equilibrium is observed in this case. (\protect\subref{fig:high_max_derivative}) shows the case where $\max_xu'(x)\geq1/\alpha, ~ (\alpha=0.6)$. Equilibria are observed in both lower and upper stable regions in this case. ($k_{min}=2, k_{max}=5,s=r=20,\epsilon=0.05$)}
\label{fig:u}
\end{figure}

\item  For $u(\cdot)$ that is very steep at the point of inflection, the iteration usually converges to one of the two equilibria, one each in the upper and lower zones, depending on the initiation. (see Figure \ref{fig:u}(\subref{fig:high_max_derivative}))

\item Theorems \ref{theorem:u_1suff} and \ref{theorem:u_2suff} only provide sufficient (and not necessary) conditions for existence of equilibria in the upper and lower stable regions. And hence, even when $u_1(b+c)<b+c$, we frequently observe an equilibrium point in the upper stable region (see Figure \ref{fig:u1}). 

\begin{figure}[h!]
\centering
\begin{subfigure}{.45\textwidth}
 \includegraphics[width=\textwidth]{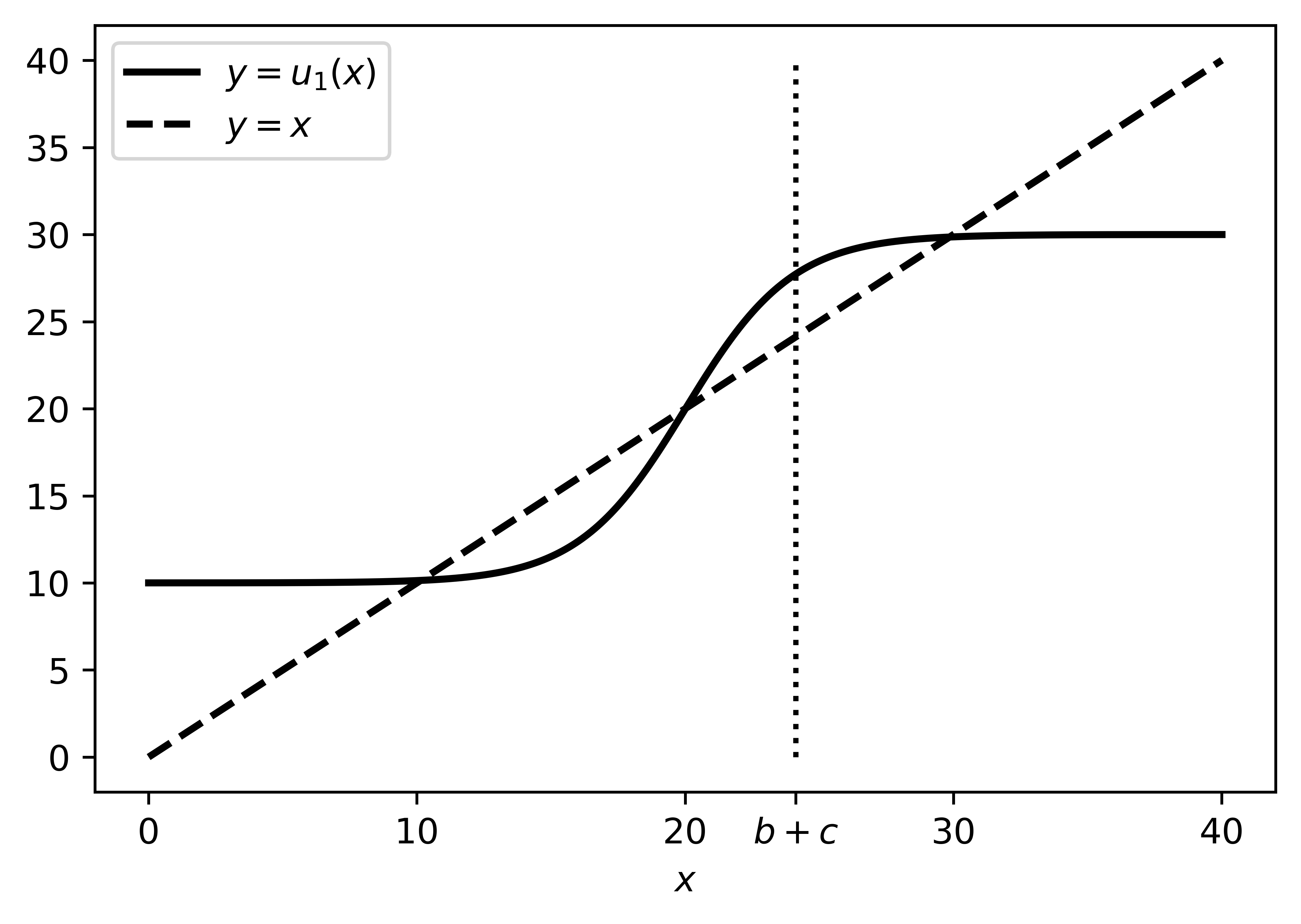}  
 \caption{}
 \label{fig:u1_greater}
\end{subfigure}
\begin{subfigure}{.45\textwidth}
 \includegraphics[width=\textwidth]{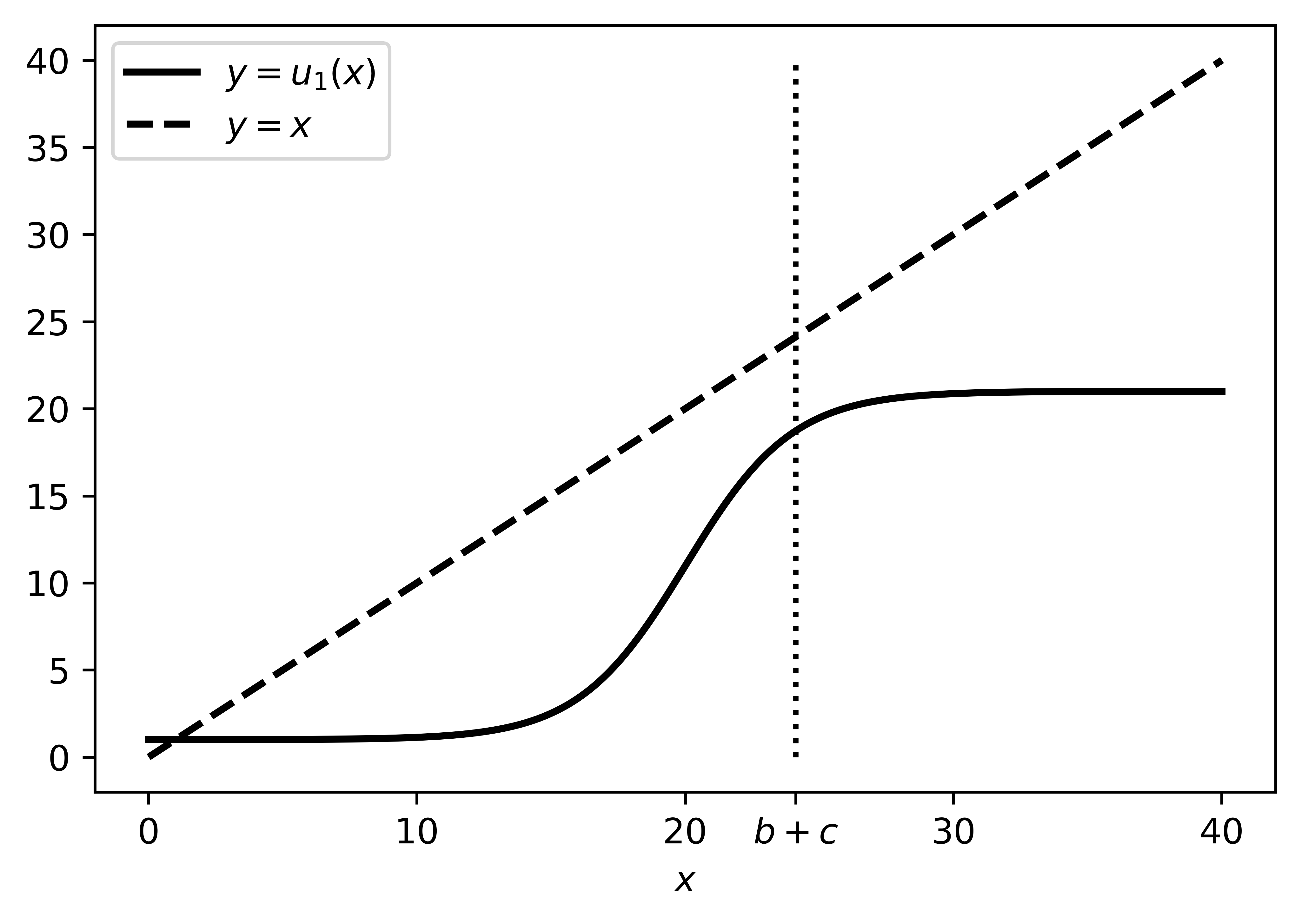}  
 \caption{}
 \label{fig:u1_smaller}
\end{subfigure}
\caption{$u_1(\cdot)$ in two different cases: (\protect\subref{fig:u1_greater}) shows the case where $u_1(b+c)\geq b+c,~ (k_{min}=10)$. (\protect\subref{fig:u1_smaller}) shows the case where $u_1(b+c)<b+c, ~ (k_{min}=1)$. In both cases, an equilibrium exists in the upper stable region. ($\alpha=0.5,k_{max}=20,s=r=20,\epsilon=0.05$)}
\label{fig:u1}
\end{figure}
\end{itemize}
\subsection{Third stable equilibrium point}
\begin{figure}[h!]
\begin{center}
 \includegraphics[width=0.5\columnwidth]{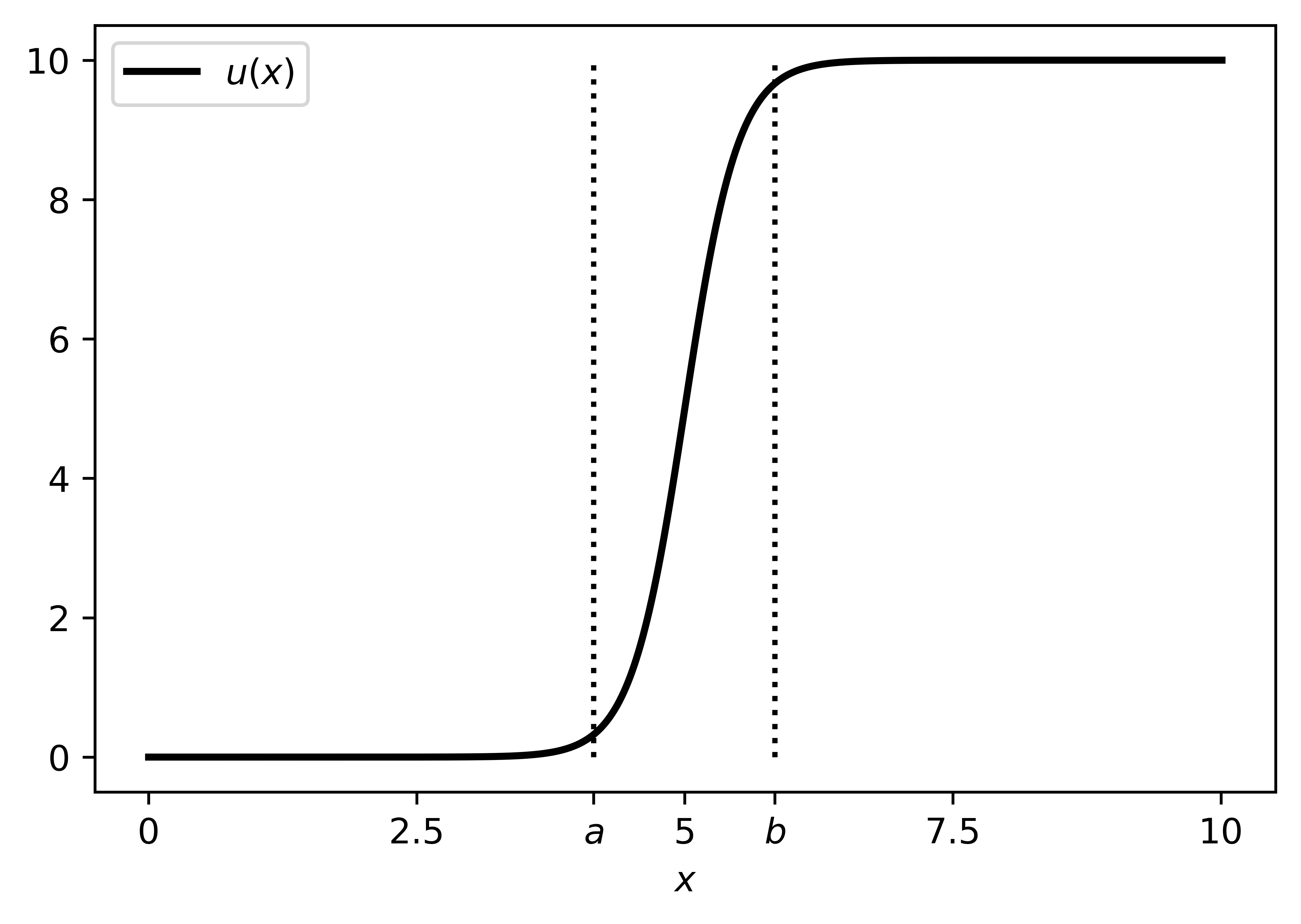}
 \caption{$u(\cdot)$ with a very steep rise constructed in order to confirm existence of a third stable equilibrium point}
 \label{fig:specific}
\end{center}
\end{figure}

In general, we observed that convergence to any equilibrium point other than the maximal and minimal ones in the upper and lower stable regions is very rare. In our initial experiments, we noticed that the iteration converged either to the maximal or to the minimal equilibrium point. To confirm that the existence of a third stable equilibrium point is possible, we manually constructed and computed the equilibrium points for a small system $(s=4,r=2)$. For further simplicity, we kept all rewards very close to $2$ (i.e., $k(i,v)\in[1.99,2.01],\forall i,v$) and fixed the value of $\alpha$ to $0.8$. We chose a very steep $u(\cdot)$ such that $u(x)\approx0,\forall x\in[k_{min},a)$ and $u(x)\approx K=10,\forall x\in(b,K]$. The function $u(\cdot)$  rises from $0$ to $10$ almost entirely between $a$ and $b$ (see Figure \ref{fig:specific}). This also ensured that the conditions in Theorems \ref{theorem:u_1suff} and \ref{theorem:u_2suff} are satisfied. The transition matrix was fixed so that for each state, the two actions are identical (i.e. $p(j|i,v)=p(j|i,w), \forall i,j$ where $w,v$ are the two actions for state $i$). For the transition probabilities we chose, we observed that there are four stable equilibria (two equilbria in addition to the maximal and minimal equilibria). When initiated in close vicinity to these additional equilibrium points, the iteration converges to them. 

Apart from this above constructed case, we never observed the Q-learning iteration to converge to a stable equilibrium point other than the maximal or minimal. While three or more stable equilibria can exist for many systems, convergence to these points seems very infrequent in randomly generated test cases, suggesting a small domain of attraction. This is a purely empirical observation and it will be interesting to see if there is a fundamental reason for it being so. 

We also compared our scheme with classical Q-learning for the same problem parameters (figures not included). The (necessarily unique) equilibrium of the latter was empirically found to be closer to the maximal equilibrium of the former than its minimal equilibrium, and either above or below it depending on the shape of the S-curve. For example, the equilibrium of classical Q-learning would be above the maximal equilibrium of our scheme if the S-curve had a very limited range. It became closer to the minimal equilibrium if we replaced reward maximization by cost minimization. Though the latter is merely a mathematical curiosity, it suggests that the preceding observations have more to do with ours being a maximization problem.

\section{Alternative Formulation}\label{sec:alt}
In our original formulation, only the future returns are distorted using the prospect theoretic valuation map. We studied another formulation where the S-shaped curve $u(\cdot)$ is applied to the total returns, i.e., both the current rewards and the future returns are distorted. 
\subsection{Q-learning scheme and its convergence}
The Q-learning iteration in this case is the following:
\begin{eqnarray}
Q_{n+1}(i,v) &=& Q_n(i,v) + a(\mu(i,v,n))I\{X_n = i, U_n = v\}\times \nonumber \\
&&\Bigg(u\Big(k(i,v) + \alpha (Q_n(X_{n+1}, U_{n+1}) - \nonumber \\
&& \xi_n(X_{n+1}, U_{n+1}))\Big) - Q_n(i,v)\Bigg). 
\label{QL_alt}
\end{eqnarray}
In this case, $u:[0,K+\alpha c]\mapsto[0,K]$ and similar to Lemma \ref{iteration-bounded}, when initiated in the set $\SA_1:=[0,K]^{sr}$, the Q-learning iteration stays in the set $\SA_1$. 

Iteration (\ref{QL_alt}) tracks to the following o.d.e.:
\begin{eqnarray}\label{ode_alt}
\lefteqn{\frac{d}{dt}q_t(i,v) = h_{1_{i,v}}(q_t)} \nonumber \\
&=& F_{1_{i,v}}(q_t) - q_t(i,v)  \nonumber \\
&:=&\int_{\R^{sr}} \Bigg\lbrack\Bigg(\sum_jp(j|i,v)\bigg((1-\epsilon)u\Big(k(i,v) \nonumber \\
&&+ \ \alpha\max_w\big(q_t(j,w) - y_{j,w}\big)\Big) \nonumber \\ 
&+& \frac{\epsilon}{r-1}\sum_{w\neq w^*_{q_t,y,j}}u\Big(k(i,v)+\alpha\big(q_t(j,w) - y_{j,w}\big)\Big) \bigg) \Bigg) \nonumber\\
&\times&\prod_{j,w} \varphi(y_{j,w})dy_{j,w}\Bigg\rbrack - q_t(i,v). 
\end{eqnarray}

As in Lemma \ref{ode-bounded} above, o.d.e.\ (\ref{ode_alt}) remains in the set $\SA_1$ when initiated in $\SA_1$. This o.d.e.\ is also a cooperative o.d.e.\ where the Jacobian matrix of $h_1$ (resp., $F_1$) at $q$ is $J_1(q) - I$ (resp., $J_1(q)$) with the $((i,v),j,w))$\textsuperscript{th} element of $J_1(q)$ being:
\begin{align}
J_1(q)_{(i,v),(j,w)}=& p(j|i, v)\alpha \nonumber\\
\times&\int\Bigg\lbrack\bigg((1-\epsilon)u'\Big(k(i,v)+\alpha\big(q(j,w) - y_{j,w}\big)\Big)I_\beta(j,w) \nonumber \\
+&  \frac{\epsilon}{r-1}u'\Big(k(i,v)+\alpha\big(q(j,w) - y_{j,w}\big)\Big)\Big(1-I_\beta(j,w)\Big)\bigg) \nonumber\\
\times& \prod_{w} \varphi(y_{j,w})dy_{j,w}\Bigg\rbrack
\label{Jacobian_alt}
\end{align}
where $I_\beta(j,w)=I_{\{q(j,w) - y_{j,w} > q(j,w') - y_{j,w'} \ \forall \ w' \neq w\}}$. (As before, we ignore the case of multiple maximizers.)

Since o.d.e.\ (\ref{ode_alt}) is a cooperative o.d.e.\ and it maps $\SA_1$ to $\SA_1$, all theorems in Section \ref{sec:monotone} can be applied to this alternative formulation as well. Thus there exist maximal and minimal equilibria for (\ref{QL_alt}) and the order interval trichotomy holds for the equilibrium points of (\ref{QL_alt}).

\subsection{Stable regions}
When $u'(x)<1/\alpha,\forall x\in\SA_1$, then the results from \cite{ShenS} again show that there will exist only one equilibrium point in the set $\SA_1$. But when $u'(x)$ crosses $1/\alpha$ in the middle regions, we observe behavior different from our original formulation. The upper and lower stable regions do not always exist in this case. As in Section \ref{sec:stable_regions}, these stable regions are defined in the regions where the sum of each row of the Jacobian matrix $J_1(q)$ is less than $1$. This will be true when $u'(k(i,v)+\alpha(q(i,v)\pm c))<\frac{1}{\alpha},\forall~ i,v$. 

So, the upper stable region is defined as $(b'+c,K]^{sr}$ where $b'=\frac{b-k_{min}}{\alpha}$ and will exist if the following condition holds:
\begin{condition}\label{cond:alt:upper}
\[
b'+c<K\Leftrightarrow \frac{b-k_{min}}{\alpha}+c<K\Leftrightarrow b<k_{min}+\alpha(K-c).
\]
\end{condition}
Similarly the lower stable region is defined as $[0,a'-c)^{sr}$ where $a'=\frac{a-k_{max}}{\alpha}$ and will exist if:
\begin{condition}\label{cond:alt:lower}
\[
a'-c>0\Leftrightarrow \frac{a-k_{max}}{\alpha}-c>0\Leftrightarrow a>k_{max}+\alpha c.
\]
\end{condition}
\subsection{Numerical experiments}
To simulate this formulation, i.e., iteration (\ref{QL_alt}) and o.d.e.\ (\ref{ode_alt}), we vary parameters as in Section \ref{sec:simu}. As expected, this scheme also converges to the equilibrium points. Most trends observed for the original scheme are observed here as well. For example, when $u(\cdot)$ rises very steeply and $\alpha$ is large (greater than $\approx$ 0.7), we observe two equilibrium points, one each in the upper and lower stable regions. 

An important difference between the two schemes lies in the values of the maximal and minimal equilibrium points. Since both the current and the future returns are distorted in this scheme, when the original scheme converged to large values of $Q$, this scheme converges to even larger $Q$ values. Similarly, when the original scheme converged to small $Q$ values, this scheme converges to even smaller $Q$ values. This is observed in the simulations as well. When same parameters are set for both schemes, the maximal equilibrium point of the alternate formulation is higher than the maximal equilibrium for the original formulation.

\section{Conclusions}\label{sec:conc}
In this work we studied classical Q-learning from a prospect theoretic viewpoint, i.e., when the valuation of future returns is distorted by an S-shaped subjective map. We then present conditions under which the iteration and its limiting o.d.e.\ converge to equilibrium points. Upper and lower stable regions are defined, in each of which at most one equilibrium  exists and is stable. Additional results regarding the number and location of equilibria are also presented. We verify these results through simulations and make further comments regarding the observations. We finally study an alternative prospect theoretic scheme where both the current and future returns are distorted. A possible avenue for future work is to characterize conditions under which there are three or more equilibria and the scheme converges to an equilibrium point other than the maximal or minimal equilibrium. Also, it will be interesting to study the bifurcation phenomena that arise when the map $u(\cdot)$ is homotopically morphed from the identity map to an S-curve.

\end{document}